# The Effects of Interaction Conflicts, Levels of Automation, and Frequency of Automation on Human-Automation Trust And Acceptance

Hadi Halvachi, Ali Asghar Nazari Shirehjini, Zahra Kakavand, Niloofar Hashemi, and

Shervin Shirmohammadi

**ABSTRACT**

In the presence of interaction conflicts, user trust in automation plays an important role in accepting intelligent environments such as smart homes. In this paper, a factorial research design is employed to investigate and compare the single and joint effects of Level of Automation (LoA), Frequency of Automated responses (FoA), and Conflict Intensity (CI) on human trust and acceptance of automation in the context of smart homes. To study these effects, we conducted web-based experiments to gather data from 324 online participants who experienced the system through a 3D simulation of a smart home. The findings show that the level and frequency of automation had an impact on user trust in smart environments. Furthermore, the results demonstrate that the users' acceptance of automated smart environments decreased in the presence of automation failures and interaction conflicts.

## 1. Introduction

User trust in automation is a major factor affecting the acceptance of intelligent systems [1]–[5] As shown by numerous research works (e.g., [6]–[11]), trust is negatively affected when automation errors occur. It can be hypothesized that lower levels of automation have a smaller probability of causing wrong actions [12] because users are kept in the decision-making loop, correcting potentially wrong system decisions before any interaction happens.

The main contribution of this work is the scientific conduct of a Three-Factor Analysis of Variance to study trust factors within the context of a simulated smart home with imperfect automation. To the best of our knowledge, this research is the first that has conducted a Three-Way Analysis of Variance (ANOVA) experiment to study the effect of the Level of Automation, Frequency of Automation, and Conflicts on user trust and acceptance of automation in the smart home context. In this regard, this work advances the field of knowledge by providing insights and delivering empirically supported answers to the hypotheses (listed in Table 3).

Using a simulated smart home and online questionnaires, we conducted experiments with 324 participants. The results of our three-way ANOVA performed on the questionnaire responses indicate that the main and interaction effects of levels of automation and interaction conflict on user's trust and acceptance of smart homes are significant. In particular, we found that in the presence of interaction conflicts, higher levels of automation decrease trust and acceptance of smart environments more than lower levels of automation. This work extends our previous works [13]–[15].

The remainder of this paper is organized as follows. Section 2 describes the general concepts of the experiment and reviews related studies. Sections 3 and 4 describe our hypotheses, 3D simulator, and research design—sections 4.6 present the research results, while section 6 provides discussion f the results and explores the limitations of the study. Finally, the paper finishes with Section 7 in which the conclusion and the suggested future work of the paper have been presented.

## 2. Literature Review

This section discusses the general concepts of the experiment, which are Level of Automation (LoA), Frequency of Automation (FoA), Conflict Intensity (CI), trust, and Acceptance. Additionally, we surveyed the related studies in the following fields: methods for studying user trust and acceptance in the specific context of smart environments involving simulators and questionnaires, the use of online experiments and crowdsourcing for human factor studies, and the use of virtual environments for studying human interaction with smart environments and systems. Table 1 presents a comparison of these prior studies.

Table 1 Literature Study

| Studies | Year | Independent Variables | Dependent Variables | Domain | Mental Model Development Tools | Experiment Method | Dependent Variables Measured by |
|---|---|---|---|---|---|---|---|
| Antifakos et al. [16] | 2005 | Displaying System Confidence | User Trust | Context-aware Mobile Phone | Video | Laboratory | System choice verification rate by user |
| Hossein et al. [17] | 2013 | Dynamic Adjustment of LoA | User Trust | Ambient-aware environment | Experimental Smart Home | Laboratory | "System Trust Scale" [18] |
| De Vries et al. [19] | 2003 | Automation Error | User Trust | Web | Online Website | Online | Questionnaire |
| Liu et al. [20] | 2012 | Testing platform | Website Usability | Web | School Website | Online | Successfully performed tasks + 6 Questions |
| Shin et al. [21] | 2008 | Conflict Resolution Method | User Preference | Context-aware Application | Smart Home 3D Simulator | Laboratory | Satisfaction Questionnaire |
| Meerbeek et al. [22] | 2016 | Level of Automation | User Satisfaction | Building Automation | Real System | Laboratory | Questionnaire |
| Haspiel et al. [23] | 2018 | LoA and Time of Explanations | User Trust | Automated Vehicles | Simulator | Laboratory | Questionnaire |
| Hartwich et al. [24] | 2018 | LoA and Age | User Acceptance | Automated Vehicles | Simulator | Laboratory | Questionnaire |
| Blömacher et al. [25] | 2018 | System Description Level | Trust and Acceptance | Automated Vehicles | Video | Laboratory | Questionnaire |
| Shin, D. [26] | 2021 | Explainability and Causability | Perception, Trust, and Acceptance | AI systems | Prior experience | Mixed | Questionnaire |
| Molnar et al. [27] | 2018 | LoA (Control preference) | Trust and Acceptance | Automated Vehicles | Simulator | Laboratory | Measurement, Interview, and Questionnaire |
| Hartwich et al. [28] | 2019 | System experience | Trust and Acceptance | Highly Automated Driving (HAD) | Simulator | Laboratory | Questionnaire |
| Clement et al. [29] | 2022 | System experience and Demographic | Trust and Acceptance | Automated Vehicles | Simulator | Not mentioned | Questionnaire |
| Schomakers et al. [30] | 2021 | LoA and application field | Privacy and Trust | Smart home | - | online | Questionnaire |

## 2.1. Level of Automation (LoA)

Level of automation refers to the level of task autonomy maintained between human operators and machines in smart systems. Depending on the degree of automation applied to each aspect of a task, levels of automation range from complete manual control to full automation. Two commonly used 10-level models of automation have been proposed by Kaber et al. [31] and Sheridan [12]. Furthermore, Hossain et al. [32] suggested a model with four levels for automation in intelligent environments. In level A, tasks are performed automatically, and the human operator has little control over the interaction. In level B, the smart system provides the user with a list of recommended tasks and services. At this level, the user can choose or modify the suggestion through interaction with the smart system. Recommendation systems like [33] can be considered as this type of interaction. In level C, the smart automation output shows information about the user's intentions or possible actions that can be taken. In level D, null action, the smart system doesn't do anything, and control is completely manual. In this research, we have used this level of automation taxonomy to design a smart home simulator.

## 2.2. Frequency of Automation (FoA)

The frequency of automation is the rate at which the automated system is programmed to respond to the user, for example, in a smart system that is designed to inform a mother whenever her newborn baby cries, the response frequency is rather high. Meanwhile, in a smart system that is designed to diagnose cancer, response frequency is low, as "having cancer" is not something that frequently happens to a person. In this research, we use two degrees of automation response frequency for designing our smart environment: high and low [34].

## 2.3. Conflict Intensity (CI)

In the literature, conflict is considered as the impossibility of an agent or group of agents reaching a goal that matters [35]. Interaction conflict refers to the situation where automation executes an action that contradicts some recent actions that were explicitly performed or requested by a human agent, e.g., turning off a light turned on by the user to reduce energy consumption [36][37]. The concept of interaction conflict is illustrated in Figure 1, where $P_n$ represents the effect of an interaction on the n-th parameter within the $T_1$ to $T_2$ time interval. The user performs an action in $T_1$ to $T_2$ time interval and causes changes in parameters $P_1$ to $P_n$ in the environment; however, the intelligent system strikes out and changes some parameters in a way that is in contrast with the user's need in that specific time interval.

The intensity of conflict is also important. Conflict intensity determines how much the automation behavior contradicts with the user's goal. There are two levels of conflict intensity: High and Low. Intensity is high when the automation acts the exact opposite of the user's command and is low when it sets an environmental parameter outside of the user's desired range [14].

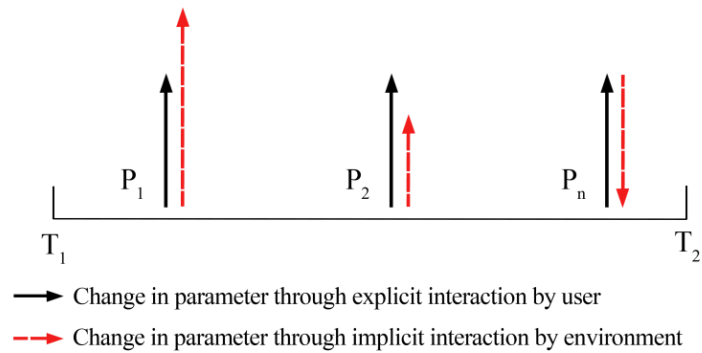

Figure 1 Interaction conflict concept in Ambient Intelligence (Martin & Shirehjini, 2008).

## 2.4. Trust

In [38], human-computer trust is defined as "the extent to which a user is confident in and willing to act on the basis of the recommendations, actions, and decisions of an artificially intelligent decision aid." Trust is one of the most important and substantial determinants of technology acceptance [9], [39], [40]. In general, there are three different types of human-system trust [16]. First, *initial trust* is the trust before any experience with a system. Second, *short-term* trust is built up by the first interaction with the system. Third, *long-term* trust is developed by long-term use of the system. The type of trust we studied in this work is short-term trust, which concerns with the trust that can be developed after the first interaction between the user and the system.

Similar to human-computer trust, human-automation trust has been defined as "the extent to which a user is confident in and willing to rely on an automated system in the presence of risks" [8], [38]. Furthermore, many researchers (e.g., [8], [9], [39]) have described that the dynamic nature of trust depends on different factors, such as users' previous experience, the quality of experience, the demography of interacting agent, and the context of interaction. Research conducted by De Vries et al. [19] on the effect of automation failures on trust in automated route planning assistance showed that a high automation error rate decreased the level of user trust. Also, Palvia [41] stated that perceived risk and perceived level of control shaped the beliefs and intentions of consumers toward trust in e-commerce. As a result of automation failures and/or the occurrence of interaction conflicts, user trust may decrease to a point at which they would reject using the automated system to prevent facing risky situations or harmful consequences.

In a recent study, Mishler & Chen [42] examined how different types of driving automation system failures (no-failure, takeover request, system malfunction) affected drivers' trust. A simulator was used for the study. Responses of the 122 study participants showed that in the takeover-request and system-malfunction conditions, trust decreased significantly in response to automation errors. In a similar study, Lyons et al. [43] examined the effect of unexpected robot behavior on human trust. The findings of their study indicated that unexpected system behaviors harmed users' trust and trustworthiness perceptions. Furthermore, the negative effect of unexpected system behavior on user trust was smaller when the system was providing explanations about its decisions [43], indicating that the loss of trust in response to errors was smaller in lower automation levels.

Shin and Park investigated how trust is related to social issues such as transparency, fairness, and accountability and how it affects the use and adoption of algorithmic systems [44]. To study trust, they used a mixed research method by integrating imperative methods and surveys. Their results showed that fairness, accountability, and transparency had notable effects on trust.

In another study [19], the effect of automation error on system trust was investigated through an experiment in the domain of route planning. During the experiment, the participants experienced both manual and automatic modes of route planning. The number of errors varied in both modes to study users' trust in different situations. They recruited 96 participants and assigned them to four experimental conditions. The experiment was conducted on the computers at the laboratory. After completion, participants were asked to rate their trust in the automatic system via a 7-point Likert scale. The results showed that a high automation error rate decreases the level of system trust.

Likewise, Haspiel et al. [23] investigated the effect of level and time of explanation about automated vehicles on user trust. This experiment was conducted in a high-fidelity vehicle simulator and employed a within-subjects design with four driving conditions. After each driving condition, participants completed a questionnaire measuring their trust. Also, in another study in the domain of the automated vehicle, Zhang et al. [45] proposed a theoretical framework that explains the relationships between individual differences, expectations, trust, and the acceptance of autonomous vehicles and empirically examined the framework. They conducted a 2x2 factorial within-subject experiment with four conditions representing two types of driving behavior (normal vs. aggressive) and two kinds of driving weather conditions (sunny vs. snowy). An appropriate mental model was developed using videos, and data collection was conducted through an online survey.

All the above studies were conducted in laboratories with 14-96 students and, at most, four experimental conditions. One challenge of group experiments with more than ten experimental conditions is the need for a large number of participants. Moreover, in many research designs, participants cannot be tested concurrently. This challenge makes two or three-factor ANOVA experiments highly time-consuming, which motivated us to use an online experiment.

### 2.5. Acceptance

Technology acceptance determines whether or not a user is willing to use the system. Technology Acceptance Model (TAM) is one of the theoretical models to specify user acceptance of information technology. TAM, originally proposed by Davis [46], is a theoretical model that helps explain and predict user behavior toward information technology. It provides a basis by which we can study how external variables influence users' beliefs about the system and their willingness to use the system. Two cognitive beliefs that TAM points to are perceived usefulness and perceived ease of use. According to TAM, a user's decision about using a system is influenced directly or indirectly by the user's behavioral intentions, attitude, as well as perceived usefulness and ease of use of the system. Further, TAM states that external factors affect intention and actual use through mediated effects on perceived usefulness and perceived ease of use.

In a study by Ziefle et al. [2], participants rated their fears about illegal access, data transfer without consent, and data loss due to technical malfunctions for medical technology at home. The fear that data could be altered or deleted as a result of system malfunctions, like sensor errors, was most pronounced. The results showed that technical disturbances played a critical role in users' acceptance of medical technologies at home.

A study by Ghazizadeh et al. in [39] extended Technology Acceptance Model and added trust as one of the main determinants and factors of their trust levels. Likewise, Khattab et al. [47] showed that the perceived ease of use, perceived usefulness, perceived risk, and trust were the four main factors affecting citizens' acceptance of government e-services.

Hossain et al. [48] carried out a study that investigated the users' acceptance of a data synchronization model during the execution of conflicting updates in an e-health environment. The data synchronization model performed in three levels of automation, namely, a) auto synchronization, b) semi-automatic synchronization, and c) user-involved synchronization. The results showed that the users' perceived acceptability of the system was significantly higher in user-involved synchronization compared to the other two levels of automation.

In a study in the domain of automated vehicles, Blömacher et al. [25] studied how system descriptions influenced drivers' acceptance. In their experiment, the participants watched five video clips with different situations. Further, acceptance before and after watching videos was measured using a questionnaire with a 5-point Likert scale developed by Van Der Laan et al. [49]. In another research, Noblet et al. [50] tested the role of varying levels of automation on the acceptance of connected autonomous vehicles using an online survey. Also, Hartwich et al. [24] examined the effect of the level of automation (two levels: manual, automated) and age on the acceptance in a driving simulator experiment. To assess the acceptance, the authors used a questionnaire presented in the study of Van Der Laan et al. [49]. The result showed that the drivers' acceptance of automated driving significantly increased after experiencing the technology for the first time.

## 2.6. Trust and Acceptance

The *critical trust level,* which is used in this paper, is a point or interval beyond which users refuse system-aided automation. Marsh et al. [51] stated this threshold of trust as the border between "Trust" and "Untrust." Figure 2 illustrates this interrelation between various trust segments and cooperation willingness. The result of the study conducted by Marsh & Dibben [51] indicated that in some situations, although trust may still be present and take some positive value, it might not be sufficient to initiate or encourage cooperation between users and the automated commerce system. Untrust is positive trust but not enough to cooperate [51]. Tan et al. [52] showed that a person only engages in an E-Commerce transaction if the level of his/her trust exceeds a threshold. Consistently, in an experiment conducted by Verberne et al. [53] the human-automation trust in adaptive cruise control systems correlated with technology acceptance. Similarly, numerous previous studies (e.g., [9], [54]–[57]) showed a significant correlation between user trust and technology acceptance, allowing for assigning participants into two disjunctive groups of "users" or "refusers" of the system based on their level of acceptance.

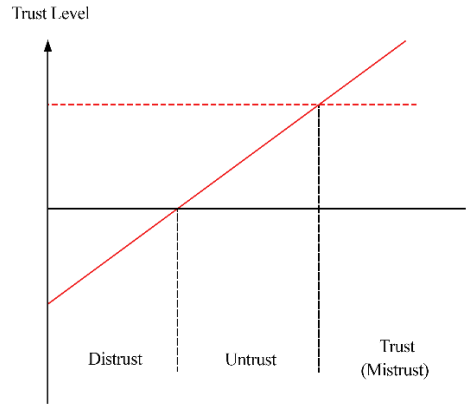

Figure 2 From Distrust to Trust [51]. Note: we reproduced this figure from [51] without written consent

## 2.7. Using online experiments and crowd-sourcing in human factors studies

Crowdsourcing is defined as the recruitment of an online, independent global workforce to work on a specifically defined task or set of tasks [58]. Crowdsourcing and online surveys help researchers to recruit large numbers of participants at a relatively low cost with fewer physical resources. Factors such as the costs associated with participants recruiting, observers and equipment acquiring, and time constraints have led to the collection of data from users via online survey tools (e.g., surveymonkey.com) [59].

Previous studies utilized crowdsourcing to acquire user information promptly. For example, Kittur et al. [59] studied the use of micro-task markets, such as Amazon's Mechanical Turk, for collecting user measurements. This method was proposed to overcome various challenges, such as monetary and time costs as well as the small number of participants. In their experiment, users were asked to rate Wikipedia articles, and then their ratings were compared with a previous experiment. In the first 24 hours, 93 ratings were received. The positive correlation between Mechanical Turk ratings and previous ones was significant.

In another study, Behrend et al. [58] concluded that the reliability of data from the crowdsourcing sample is equally good or better than that of the university sample. They collected both traditional university pool samples and Mechanical Turk samples to conduct a survey. The crowdsourced samples were more diverse in demographic characteristics and were higher in social desirability. Moreover, internal consistency was higher in the crowdsourced samples.

Crowdsourcing was also used as a testing method to study the usability of applications and websites [60]. Liu et al. [20] conducted two similar studies to evaluate a website; One study via a traditional lab setting and the other through crowdsourcing. After analyzing the results of both studies, they concluded that the identified usability problems of the website overlapped significantly.

Although online experiments allow for recruiting a larger number of participants, yet, there are some challenges and limitations associated with online experiments. The possibility of occurrence of distraction during the experiment, difficulties in verifying the identity of subjects, and the possibility of careless reading of instructions are the main drawbacks of online experiments [19].

## 2.8. Studies using virtual reality-based simulators

Conducting real-world experiments involves challenging tasks, including acquiring expensive appliances, collecting information from sensors, and time and monetary constraints. Moreover, real-world experiments require a running prototype. Where real-world experiments are not feasible, researchers use virtual environments and online surveys for human factor studies. However, a drawback of online questionnaires when applied to smart environments is that respondents may lack the mental models that are necessary to provide adequate responses. To address this drawback, many researchers used simulators or virtual environments in the field of human-computer studies (e.g., [11], [21], [61] [62]. Simulators were also used to study operator trust in and use of automated systems (e.g., [11], [21], [61], [63]–[68].

In a previous work [15], we studied whether the use of virtual reality-based simulations supported the development of mental models. In particular, we used a web-based Interactive Realistic Virtual Reality based Simulator (IRVRS) to investigate how its use affected the mental models of virtual smart home users. Think Aloud and Card Sorting methods were used to assess the development of the participant's mental models. The users were asked to sort the cards before and after interacting with the 3D environment. The results indicated that a short-time experience with Interactive Realistic Virtual Reality based Simulations had a positive effect on the mental models of inexperienced users.

In another previous work [69], we combined crowdsourcing and a web-based IRVRS for a human-automation trust study. Using the web-based IRVRS, participants could interact with a simulated smart home, allowing participants to gain initial experience with the automated system under study. Our findings showed that the use of IRVRS supported users in developing an initial mental model of the system under study [13], [14]. Compared to the users without any mental model, the participants who had used the simulation were able to provide more accurate responses.

## 3. Developing research hypotheses

This research investigated the effects of the level of automation (i.e., 0, 1, 2, 3), frequency of automation (i.e., low or high), and conflict intensity (i.e., low or high) on users' trust and acceptance of smart environments. The dependent variables were users' trust and acceptance of automation.

### 3.1. Hypothesis 1 ($H\_1$): The effect of conflict intensity and level of automation on trust and acceptance

Trust in automation depends on the level of control the operator has over automated functions [7], [11], [53]. At lower levels of automation, users are asked to confirm a decision or choose from a relevant set of actions, increasing the perceived user control [12], [70]. When perceived control increases, trust in automation also increases [7], [11], [53].

It has been shown that automation that performs actions while providing information to the operator (~Level 7 automation in [12]) is more trustworthy than automation without providing any information to the operator (~Level 10 automation in [12]) [53].

It is easier for the user to trust an automated system when there is more control in the user's hands [7], [11], [53].

De Visser et al. [7] assessed user trust in a human-robot team in three conditions (a) Manual, (b) Static (fixed level of automation), and (c) Adaptive Automation (dynamic level of automation based on mission context) [7]. The results of their experiment showed that the level of trust in automation was higher in cases where the level of automation was adapted to the context.

Meerbeek et al. [22] investigated the effect of the LoA and expressiveness of an automated blinds system installed on a virtual window on users' satisfaction and acceptance [22]. Their results demonstrated the potential of choosing the right LoA and expressive interfaces to increase user acceptance.

Suppose three different scenarios as listed in Table 2Table 2. An agent provides automated aid with purchasing food. Assume the user has an important work session. The user wants to invite the participants to lunch before the session. We consider the smart agents' behavior for three levels of automation.

Table 2 Comparison of different levels of automation

| Level of Automation | Description | Effects of Conflicts | | |
|---|---|---|---|---|
| | | Time | Cost | Recoverability |
| LoA_1 | System detects the goal of user (work lunch) and suggests a menu. | < 10 second | 0 | yes |
| LoA_2 | System detects the goal of user, offers 10 Cheeseburgers, and informs the user. | 2 min | 50$ | Forward Recoverable |
| LoA_3 | System detects the goal of user and offers 10 Cheeseburgers. | 2 min | Reputation Damage Failed Work Session | No |

As illustrated in Table 2, automation errors in different levels of automation will cause different costs, time loss, and effort for recovery. Increased costs and wasted time will increase the consequences of errors on the user. Consequently, perceived risk and the level of trust will be affected. For example, the conflict in the lowest level of automation (LoA_1: inappropriate lunch menu) is recoverable (user can reject suggested menu and change it) and has less impact on the trust level in comparison with conflict in higher levels of automation (such as LoA_3). The conflict in LoA_3 is not recoverable and could have great consequences on the result of the work session.

Conflicting/error-prone automation can be corrected or avoided at lower automation levels, whereas at the highest automation level (i.e., *Full Automation,* which is defined in [70]), "the system carries out all actions and the human is completely out of the control loop and cannot intervene." While the effect of the level of automation and interaction conflict on trust and acceptance has not been studied before in the context of smart homes, it is intuitive to make the following hypothesis:

*H_1: Interaction conflict's intensity has a greater negative effect on the level of Human Automation Trust (HAT) and user acceptance when it occurs in cases with higher levels of automation (LoA).*

Hypothesis H_1 states that the way conflicts influence HAT and acceptance depends on the level of automation, i.e., whether or not the human was in the decision-making loop.

## 3.2. Hypothesis 2 (H_2): The effect of conflict intensity and frequency of automation on trust and acceptance

The frequency of automation is a determinant of user trust and acceptance. Very little research has been conducted to investigate the effect of FoA on human factors such as trust and acceptance. For instance, Wilhite and Ling [71] investigated the effect of the frequency of feedback on users' understanding and satisfaction with a billing system in the energy consumption domain [71]. In another study, Funk et al. [72] mentioned that adjusting FoA appropriately increases production efficiency and users' happiness and satisfaction. The above works indicate a relationship between FoA and the users' satisfaction and acceptance. Based on these, we hypothesize that in the presence of conflicts, a higher FoA leads to more loss of trust and acceptance. Therefore, we state the following hypothesis:

*H_2: With a higher frequency of automation, the occurrence of an interaction conflict will result in a greater loss of trust in and acceptance of automation.*

For each hypothesis, Table 3 presents dependent variables, independent variables, and a list of sample research that state the same hypothesis in other domains and situations.

Table 3 Summary of our hypotheses

| ID | Hypothesis | Dependent Variables | Independent Variables | Sample Researches |
|---|---|---|---|---|
| H_1 | Interaction conflicts have greater negative effect on the level of human automation trust and user acceptance when they occur in cases with higher level of automation. | Trust, Acceptance | Conflict Intensity, Level of Automation | [7], [11], [32], [53], [83] |
| H_2 | With a higher frequency of automation, the occurrence of an interaction conflict will result in a greater loss of trust | Trust, Acceptance | Frequency of Automation | [71], [72], [84] |

## 4. Methodology

Our research method was comprised of nine main steps, as shown in Figure 3. For each experimental situation, we defined a home automation scenario. We used a 3D virtual environment system to simulate these scenarios. Using interactive 3D simulations helps users to develop appropriate mental models about the system under study [13], [14].
Users evaluated their trust in the system using a web-based questionnaire [18]. After the online data gathering, the data cleaning process started, which removed overt bias or invalid responses. Finally, hypotheses were tested using appropriate statistical analysis methods. In the remainder of this section, we explain each step of our research method, as illustrated in Figure 3.

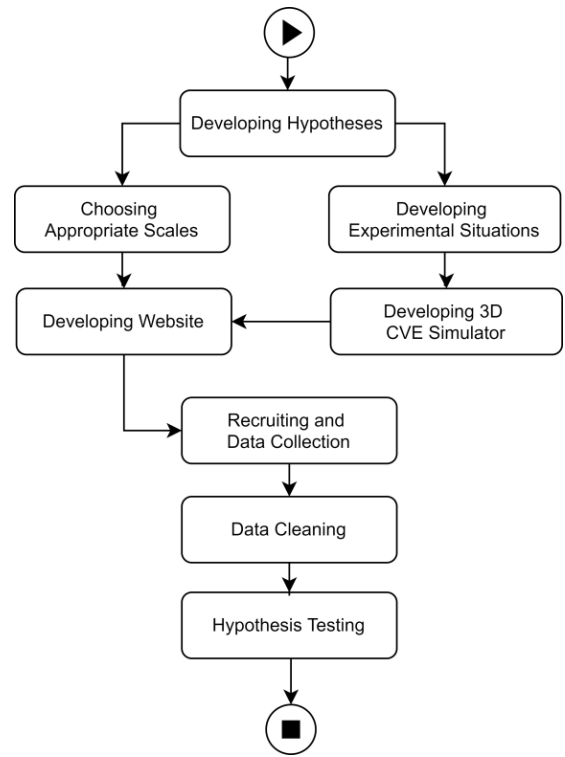

Figure 3 The proposed research method

### 4.1. Choosing Appropriate Scales

#### 4.1.1. Trust

As stated by [16], there is no "way" to measure trust directly. However, many scales for indirect trust measurement exist [18], [73], [74]. It should be noted that each of the existing scales has been developed to measure trust in a very specific domain [73], [75].

The scale presented by Jian et al. considers human-automation trust more generally and can be utilized in all automatic systems. Additionally, we aim to measure short-term trust. The only scale that is capable of measuring short-term trust is developed by [18]. We decided to use the scale provided by [18] because of its generality.

It proposes several components for trust and contains 12 questions, each evaluating a proposed component such as familiarity, honesty, etc. All the items aggregately represent user trust.

### 4.1.2. Acceptance

The user's acceptance of the system was measured with a 7-point Likert scale presented by [53]. The used acceptance questionnaire is based on the questionnaires presented by [49], [76].

### 4.1.3. Conflict Intensity Levels

To reconfirm the two-level model of conflict intensity, we conducted a separate experiment. We defined 13 conflict types and asked 11 participants to rate each conflict type on a 5-point Likert scale. Point 1 is the weakest, and point 5 is the strongest. Eleven participants rated the conflict types. The following two examples illustrate the various types of conflicts:

- Bob turns on the lights in his backyard, while the smart environment turns them off [because they consume more energy than the system thinks is appropriate for lighting the backyard.]
- Alice needs a siesta, so she turns off the lights, but the smart environment rolls up the blinds.

We conducted the Repeated Measure ANOVA test. The results indicated a statistically significant effect of conflict type on conflict intensity, $F(12, 130) = 5.2573$, $p < 0.0001$. As a precondition of the Repeated Measure ANOVA, the homogeneity of variances of residuals of the different groups was tested using the Bartlett test. No significant difference in variances was found (K-squared = 8.9718, p-value = 0.7053), indicating the required homogeneity. Next, we applied Tukey's Post Hoc test to determine which of the conflict types had a significantly different intensity. As a result, conflicts were classified into two groups: weak conflicts and strong conflicts. Strong conflicts occur when for example, the user explicitly turns something off or on, but the environment performs the exact opposite of the user's command (e.g., the first example that is mentioned above this paragraph). Weak conflicts occur when the user changes an environmental parameter, but the environment sets that parameter outside of the user's desired range but not with the exact opposite action. For example, the user turns the light off, but the environment rolls up the blinds.

### 4.2. Developing Experimental Situations

The number of independent variables and the range of their expected values determine how many experimental situations need to be considered in an experiment. In our experiment design, the smart home system can act on four different levels of automation. Also, both frequencies of automation and conflict intensity had two levels in our experiment, resulting in 16 (4*2*2) situations in total.

In the next step, we designed scenarios based on our experimental situations. In each scenario, users conducted 3-5 smart home control tasks within a designed web-based smart home simulator, such as turning on the lights, opening a door or window, etc.

From the level of automation point of view, we designed four types of scenarios: scenarios with LoA = 0…3. In scenarios with LoA = 0, the smart home only executes the user's explicit commands without making any decisions or changing the environment. In scenarios with LoA = 1, the smart home analyzes the context and sensor data and only provides some information or suggestions to the user. For example, when the energy consumption is high, the smart home system shows a notification to the user about the high energy consumption and its costs. In scenarios with LoA = 2, the smart home decides to perform some actions depending on the context, conditions, and some defined rules, but it needs the user's confirmation to execute the action. In such scenarios, the user can accept the system-suggested action (by clicking on the "Yes" button) or reject it. In scenarios with LoA = 3, the smart home system executes the actions it wants. The user is not only unable to prevent the execution of system actions but also lacks being informed about the system's performed actions explicitly.

From the frequency of automation point of view, we have two scenario types: scenarios with low frequency and scenarios with high frequency of automation. In scenarios with low FoA, notifications, confirmation requests, and actions occur one or two times in the scenario session. These occurrences increase up to five or more times in scenarios with high FoA. For example, in a high FoA scenario with LoA = 1, an energy consumption warning notification appeared to the user every 30 seconds.

From the conflict intensity point of view, we designed two types of scenarios: scenarios with low conflict densities, which contain no conflict or one weak conflict in the scenario, and scenarios with high conflict intensity, which contain more than three conflicts. Below, an example of a high automation level (~Level 3 automation), low frequency of automation responses, and high conflict intensity scenario is explained. This scenario contains two strong conflicts.

*It is approximately 4 pm on a summer day when the user comes back home from work. By approaching the entrance door, the door will be opened to the user, who is identified as the owner of the house. By entering the house, the blinds are pulled up, and the air conditioner is turned on automatically. The user attempts to turn on the lights through the user interface, but conflict occurs, and the smart home decides to turn off the lights to save energy. The user repeats his/her effort to turn the lights on, but the smart home does not allow him/her to because of the high level of automation. Next, the user turns on the TV through the user interface. When the user goes out, all the automatically turned-on devices will be turned off.*

### 4.3. Developing the Simulator and Website

A first version of the simulator was reported in our previous works [13], [14]. To conduct the current study, the smart environment simulator was changed to support the current factorial experiment. Further, to reduce bias in responses that could result from our experimental system's usability errors, we conducted intensive usability testing to discover usability problems. Following those usability testing results, we iteratively improved the system.

The simulator provided a virtual living room containing a TV, a sofa, an air-conditioner, a fan, a lamp, two windows with shields, and a door entrance. Each participant could navigate within the simulated house using arrow keys as a simple navigation tool.

The simulator was generated using the Unity game engine. Unity is a 3D game engine that supports different platforms (e.g., WebGL) and can transfer the content to the web browser. The Android app is available at [77].

WebGL is a powerful API that is incorporated into Firefox, Chrome, etc., and provides 3D content on the web [78] without application installation or additional components. Figure 4 is a screenshot of the developed simulator. Controllable devices were shown to users with red arrows.

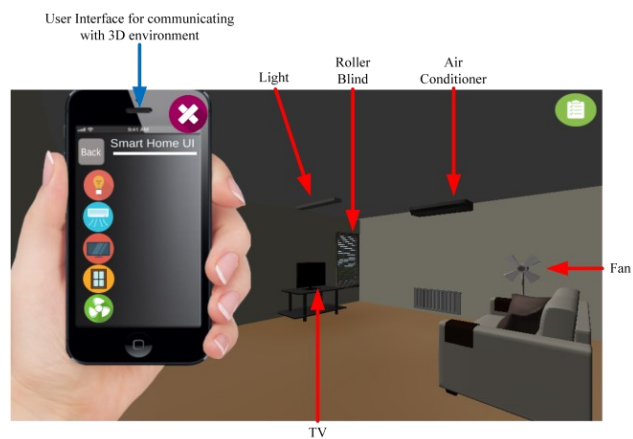

Figure 4 Screenshot of 3D CVE developed by unity game engine, showing virtual user interface and controllable devices in the virtual environment

Java script was used to implement an online questionnaire system. Responses were stored in a MySQL database. To prevent invalid submissions, embedded mechanisms such as Java scripts were conducted to validate the range and type of user inputs before storing the responses.

Context awareness was supported in our simulator, as illustrated in the following example. For instance, in some scenarios, when the user enters the house, the lights would turn on because of his/her presence.

Figure 4 shows a screenshot of the simulator. A mobile phone UI is presented on the upper left corner of the screen, through which the user can control the house. The phone has a menu with icons representing the air conditioner, the lamp, the TV, the window shields, and the fan. By clicking on each of these icons, the user can turn the representative device on and off, or in the case of window shields, the user can open or close them. After turning each device on or off, the user can instantly see the effect of their task in the 3D environment. For example, if the user closes the window shields through the phone UI, they can instantly see the window shields closing in the environment, and they can also hear the sound of shields closing.

The house communicates with the user through messages that are represented in dialogue boxes at the bottom of the screen. The frequency of message appearances in each scenario determines the level of response frequency.

Through each scenario, the user is given a list of three tasks, which is illustrated in the upper right corner of the screen. Each task is comprised of turning a device in the house on or off. The user has to accomplish these tasks by controlling the house with the given phone UI. After each task is done, a checkmark appears beside it on the task list. In the first scenario, after the user accomplishes all of the tasks on the list, they are redirected to the second scenario, where they have to do another set of three tasks. After finishing both scenarios, the user is asked to enter the survey page.

### 4.4. Recruiting and data collection

We conducted an intensive recruitment strategy using online channels. A total of 17 colleagues and friends helped us distribute the recruitment message on social media and HCI mailing lists, such as SIGCHI. In addition, the experiment link was shared on our social networks on Facebook, Twitter, LinkedIn, and ResearchGate. Thanks to the Snowball Sampling method [79], our posts were viewed over 7000 times on social media. We estimate to have reached over 1000 people by email. By the end of three weeks of recruiting time, we had 324 response data. Table 4 summarizes the distribution of responses and conversion rates based on device type (PC/Laptop/Mobile).

Table 4 Distribution and conversion rate of received responses

| Device Type | Participants Number | Recruiting Method | Website Conversion Rate |
|---|---|---|---|
| Mobile | 175 | Online and Poster | 175/395 = 44% |
| PC/Laptop | 149 | Online and Poster | 149/248 = 60% |

#### 4.4.1. Questionnaire

A total of 43 questions were asked in the questionnaire. The questionnaire can be divided into three main parts: Questions relating to the demographic characteristics of the participants and their experience with technology, the trust questionnaire, and the acceptance questionnaire.

The first part of the questionnaire addressed the participant's demographic characteristics (age, gender, education level, education field, country). Regarding user-specific aspects, the users were asked to rate their computer experience and technology affinity on a 7-point Likert scale.

The trust part of our questionnaire was a set of twelve questions provided by [18]. These questions majorly relate to how the users perceive the smart home's behavior and how confident they feel towards the smart home. The last question in this set asks the user to rate their familiarity with the smart home. All the questions in this set are answered on a 7-point Likert scale.

### 4.4.2. Participants

The participants were on average 26 years old (SD = 6.1, min=18, max=54), with 83.6% males and 16.4% females. Regarding the education level of participants, 20.3% of them held a 2-year degree, 30.5% had a 4-year degree, 42.2% had a master's degree, and 7% had a Ph.D. The participants were asked to rate their computer experience and technology affinity on a Likert scale from 1 to 7. Both in computer experience (M = 5.5, SD = 1.7, min = 0, max = 7) and technology affinity (M = 5.1, SD = 1.5, min = 0, max = 7). The collected data show that the level of computer experience and technology affinity is high on average among participants. A summary of the demographic characteristics of participants is given in Table 5.

Table 5 Summary of participants' demographic characteristics

| Age | | | Gender | | | Education Level | | | Technology Affinity | | Computer Experience | |
|---|---|---|---|---|---|---|---|---|---|---|---|---|
| Range | No. | % | Gender | No. | % | Level | No | % | Level | % | Level | % |
| Under 18 | Removed | | Male | 107 | 83.6 | 2-years degree | 26 | 20.3 | 1 | 0.7 | 1 | 3.9 |
| 18-25 | 60 | 46.9 | Female | 21 | 16.4 | 4-years degree | 39 | 30.5 | 2 | 5.5 | 2 | 3.2 |
| 26-35 | 62 | 48.4 | | | | Master's | 54 | 42.2 | 3 | 3.9 | 3 | 4.7 |
| 36-45 | 4 | 3.1 | | | | Ph.D. | 9 | 7 | 4 | 14.1 | 4 | 9.5 |
| 46-55 | 2 | 1.6 | | | | | | | 5 | 28 | 5 | 16.7 |
| Above 55 | 0 | 0 | | | | | | | 6 | 22.7 | 6 | 16.9 |
| | | | | | | | | | 7 | 25.1 | 7 | 45.1 |

### 4.4.3. Procedures

The experiment was conducted over the Internet. The website used a PHP server-side scripting language and a MySQL database to store 3D contents and participant responses. The experiment could be accessed from any system with an Internet connection and Firefox version 5 or newer, Chrome version 12 or newer, Safari version 5.1 or newer, or Opera version 11 or newer. The top of the web page contained the following introduction to the 3D content:

> *"The following 3D Simulator simulates the Smart Home. You will be asked to perform some typical tasks. Please do them in the order you have been asked to (click the Note Button to check them). Then, fill out the questionnaire. Imagine you are the owner of the Smart Home, and this is your first time using this home. There are two ways of triggering tasks; some tasks will be performed automatically, and other tasks need to be triggered by you using the UI. Note you will have no control over automatically triggered tasks. Due to the intelligent conflict management feature of the Smart Home, some of your commands may not be performed if the system has classified them as conflicting interactions."*

Participants were randomly assigned to a pair of experimental scenarios that they experienced one after the other. After they managed to finish tasks in the scenarios, they were redirected to the online questionnaire form. Note that both scenarios simulated one experimental situation. Completing the questionnaire took about 10 to 15 minutes. Data were collected worldwide in the winter of 2018 from a large and diverse variety of participants. Participation was voluntary and was not financially compensated. From this online data, we removed overt bias or invalid responses, as explained below. Finally, the hypotheses were tested using statistical analysis methods (see section 4.6).

## 4.5. Data Cleaning

Several measures of a participant's survey-taking behavior can determine the quality of their responses. For example, we used the completion time of the questionnaire (as the number of minutes) to flag responses if the completion time of the survey was less than 7 minutes. Also, the consistency of responses was assessed using pairs of items that should have opposite responses: "This home is deceptive" and "This home is reliable." In the case of identical responses, both were flagged. Next, pairs of items that should have similar responses were checked: "I can trust this home" and "I think the home is reliable." Participants with more than 2 points difference in same-meaning questions were flagged to be dropped from data analysis. This automatically removed participants with the so-called Straight-Lining behavior, which occurs when a participant answers all the questions identically without reading them. The entry IP for each response was checked, and if there were more than five responses from the same IP, they were all flagged. This was done to avoid fake responses. Also, all the data from users aged under 18 were removed. In the end, all flagged responses were eliminated from data analysis.

## 4.6. Data analysis and hypothesis testing

Statistical analyses were performed on the participants' responses to trust and acceptance. To calculate an acceptance score, we performed a maximum-likelihood factor analysis with a Promax factor rotation [80] on the responses given to the acceptance questions: "Using this home would make it easier to do my routine tasks," "I would find this home useful.", "I would buy this home," "I think that I can decide any time on tasks myself," "This system leaves me sufficient control over tasks," "The home will be easy to use," "This home is a competent performer," and "The Smart Home can be counted on." Similarly, we calculated one trust and one distrust score. As the first six questions in the trust questionnaire relate to distrust, and the latter six questions relate to trust, the overall trust scores were calculated by factor analysis, giving one negative score to each distrust question and one positive score to each trust question.

Using the Shapiro-Wilk test [81], the normality of data distribution was checked for all 16 scenarios. The internal consistency, using Cronbach Alpha, was computed to be 0.91 for our sample, which is considered as "Excellent" [82]. We conducted a factorial analysis of the variance on the overall trust score, acceptance score, and also on each questionnaire item separately. This was done to study the main and joint effects of the LoA, FoA, and CI on trust in and acceptance of automation. All behavioral data were analyzed with the R programming language version 3.4.

## 5. Results

A three-way analysis of variance was conducted on the influence of three independent variables (LoA, FoA, and CI) on the users' trust. LoA included four levels (0, 1, 2, 3), and FoA and CI consisted of two levels (low and high). All main effects were statically significant at the .05 significance level except for the FoA. The main effect for the level of automation yielded an F ratio of $F(3, 112) = 5.95, p < .001$, indicating a significant difference between levels of automation. The main effect for frequency of automation yielded an F ratio of $F(1, 112) = 1.77, p = .19$, indicating that the effect for frequency of automation was not significant. The main effect for conflict intensity yielded an F ratio of $F(1, 112) = 4.28, p = .04$, indicating that the users' trust was significantly higher for low conflicts. Only the two-way interaction effect of the level of automation and conflict intensity was significant, $F(3, 112) = 7.19, p < .001$, indicating that the effect of the LoA was greater in high CI scenarios than in low conflict intensity scenarios. There was not a significant three-way interaction between LoA, FoA, and CI, $F(3, 112) = 0.48, p = .69$. The Mean and Standard Deviation of factors are presented in Table 6Table 6. Also, a three-way analysis of the variance was conducted on the influence of the three independent variables on the acceptance. These results are presented in Table 7.

Table 7 Descriptive Statistics

| Factor | Level | Trust | | Acceptance | |
|---|---|---|---|---|---|
| | | Mean | SD | Mean | SD |
| Level of Automation | 0 | 6.71 | 1.81 | 3.93 | 1.22 |
| | 1 | 5.12 | 2.31 | 3.75 | 1.25 |
| | 2 | 6.35 | 1.00 | 4.43 | 0.71 |
| | 3 | 5.83 | 1.49 | 3.89 | 1.02 |
| Frequency of Automation | Low | 6.19 | 1.51 | 3.97 | 0.99 |
| | High | 5.82 | 2.05 | 4.03 | 1.18 |
| Conflict Intensity | Low | 6.30 | 2.02 | 4.24 | 0.94 |
| | High | 5.70 | 1.52 | 3.76 | 1.18 |

Table 6 The 3-Way ANOVA Summary Table for Acceptance

| Source | Sum Sq | df | Mean Sq | F | Sig. |
|---|---|---|---|---|---|
| Level of Automation | 8.44 | 3 | 2.81 | 2.77 | 0.04 |
| Frequency of Automation | 0.08 | 1 | 0.08 | 0.07 | 0.79 |
| Conflict Intensity | 7.42 | 1 | 7.42 | 7.31 | 0.01 |
| LoA * FoA | 9.12 | 3 | 3.04 | 2.99 | 0.03 |
| LoA * CI | 8.31 | 3 | 2.77 | 2.73 | 0.047 |
| FoA * CI | 2.31 | 1 | 2.31 | 2.28 | 0.13 |
| LoA * FoA * CI | 1.74 | 3 | 0.58 | 0.57 | 0.63 |
| Residuals | 113.70 | 112 | 1.02 | | |

Taken together, these results suggest that the LoA and CI have significant effects on users' trust in and acceptance of smart environments. Specifically, our results suggest that conflicts have stronger effects on trust and acceptance for higher levels of automation. However, the results did not indicate any significant effect of the FoA on trust or acceptance. The reason could be an insufficient or inadequate perception of differences in FoA degrees in our scenarios.

**First hypothesis:** *Interaction conflict's intensity has a greater negative effect on the level of HAT and user acceptance when it occurs in cases with higher levels of automation (LoA).*

A two-way analysis of the variance was conducted on the influence of the two independent variables, LoA and CI, on the users' trust. All effects were statically significant at the .05 significance level. The main effect for the LoA yielded an F ratio of $F(3, 120) = 6.09, p < .001$, indicating a significant difference between levels of automation. The main effect for CI yielded an F ratio of $F(1, 120) = 4.47, p = .04$, indicating that the users' trust was significantly higher for low CI. The interaction effect was significant, $F(3, 120) = 7.36, p < .001$, indicating that the effect of the LoA was greater in scenarios with high CI than in scenarios with low CI.

A two-way analysis of the variance was conducted on the influence of the aforementioned two independent variables on the acceptance of smart home automation.

To show the details of the interaction between levels of automation and CI, the simple effects of these variables on trust were analyzed. Although comparisons of the trust score by CI within each group showed that there were no significant differences in trust by levels of automation, the comparisons by LoA within each group showed there were significant differences in trust with different Cis.

With low CI, there was a statistically significant difference between different levels of automation as determined by one-way ANOVA ($F(3, 60) = 12.14, p < .001$). A Tukey post hoc test revealed that the users' trust was significantly lower in LoA = 1 ($p < .001$) and LoA = 2 ($p = .02$) compared to LoA = 0. Users' trust was significantly higher in LoA = 2 ($p = .02$) and LoA = 3 ($6.71 \pm 1.05$, $p < .001$) compared with LoA = 1. There was no statistically significant difference between LoA = 0 and LoA = 3 ($p = .18$) and between LoA = 2 and LoA = 3 ($p = .78$).

Similarly, with high CI, there was a statistically significant difference between different levels of automation as determined by one-way ANOVA ($F(3, 60) = 12.8, p < .001$). A Tukey post hoc test revealed that the users' trust was significantly lower in LoA = 1 ($p < .001$) and LoA = 3 ($p = .03$) compared to LoA = 0. Users' trust was significantly higher in LoA = 2 ($p < .001$) and LoA = 3 ($p = .02$) compared to LoA = 1. There was no statistically significant difference between LoA = 2 and the levels of automation LoA = 0 ($p = .43$) and LoA = 3 ($p = .53$).

The pattern of interactions suggests that at higher levels of automation, interaction conflicts negatively affect users' trust more than at lower levels of automation. Changes in trust scores have a descending trend at higher levels of automation.

These results can be implied to design adaptive levels of automation for smart environments, to preserve users' trust in case of interaction conflicts. In scenarios where there is a risk of automation failure or other types of interaction conflict, the system should reduce the LoA and put more control in users' hands to preserve trust.

**Second hypothesis:** *With a higher FoA, the occurrence of an interaction conflict will result in a greater loss of trust.*

Two separate two-way analyses of the variance were conducted to compare the effects of the FoA and CI on users' trust in and acceptance of automation. Only the main effect of CI was significant at the .05 significance level.

The Pearson product-moment correlation coefficient was computed to assess the relationship between the trust in and acceptance of smart homes. There was a positive correlation between the two variables, $r(126) = 0.57, p < 0.001$. The scatter plot in Figure 5 summarizes the results. Overall, there was a strong, positive correlation between trust and acceptance. Decreases in trust were correlated with decreases in acceptance of smart environments. Consistently, a one-way analysis of the variance showed that the trust score has a significant effect on the acceptance score, $F(1, 126) = 60.52, p < .001$. A summary of our results is given in Table 8.

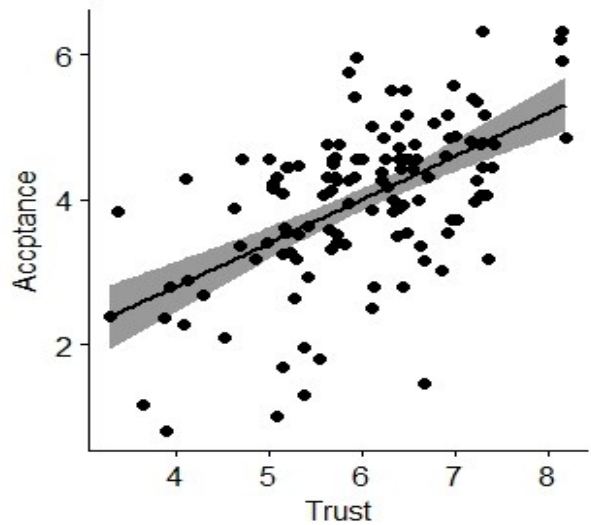

Figure 5 The relationship between trust score by acceptance score

Table 8 Summary of our results

| ID | Hypothesis | Dependent Variables | Independent Variables | Test Method | Result |
|---|---|---|---|---|---|
| H_1 | Interaction conflicts have greater negative effect on the level of human automation trust and user acceptance when they occur in cases with higher level of automation. | Trust, Acceptance | Conflict Intensity, Level of Automation | 2-Way ANOVA | Accepted |
| H_2 | With a higher frequency of automation, the occurrence of an interaction conflict will result in a greater loss of trust. | Trust, Acceptance | Conflict Intensity, Frequency of Automation | 2-Way ANOVA | Rejected |

# 6. Discussion

In general, potential users of systems can be divided into two categories: with and without mental models of the automated system. For the first category, using the simulation is not expected to change much in terms of their already existing mental model about a system under study (e.g., a smart home that provides automation at various levels). In contrast, for the latter category, without any mental models about a system under study, using a simulation will allow the development of initial mental models. Thus, using a simulation in an experiment could help in reducing subject bias for participants of the latter category, while participants from the former category, if any are available, will not be negatively affected.

This research was designed to measure short-term trust in smart environments. Although the research was carefully prepared and has reached its aims, there were some limitations. First, some participants were familiar with this technology as they were recruited from university students and HCI mailing lists. Hence, participant bias is one of the limitations of this study. Second, although the use of a 3D-based smart home simulator proved effective in supporting mental model development in a similar research [15], the results can still differ from experimentation in real settings. Finally, another limitation of this study is that the experiment could only be accessed through web browsers and Android smartphones.

It should be noted that we don't claim external validity for our findings. Rather, the findings are meant to be interpreted within the scope of our simulated application domain, user segment, and similar tasks. The external generality would need to be validated in future work [4].

# 7. Conclusion

In this study, we conducted an experiment to investigate the effects of Level of Automation (LoA), Frequency of Automation (FoA), and Interaction Conflict Intensity (CI) on Human-Automation Trust (HAT) and acceptance of smart home automation. The Investigation was done through a web-based 3D simulator and online surveys, allowing us to collect data from an extensive and diverse pool of participants. Our findings indicate a significant effect of CI and LoA on user trust and acceptance.

It is important to note that the indicators of trust and acceptance may vary in different domains. For example, if we study in the domain of automated driving instead of smart home, other indicators of trust and acceptance, such as driver's control preferences, duration of placing hands on steering wheels, or driver's focus on the road, could be the relevant measures. However, some of these indicators –like duration of placing hands on steering wheels- are not suitable for online studies. Accordingly, the 3D environment should be modified to represent the domain under study and the appropriate measures. In other words, planning scenarios according to the selected measurable indicators of trust in the target domain is vital.

In future research, we aim to experiment with an authentic smart home setting. This will help us address the mentioned shortcomings of online experiments . As we mentioned, the current experiment scenarios were designed for the smart home domain. By expanding the scope, we suggest investigating the effect of LoA, FoA, conflict intensity, and system error on user trust in and acceptance of automated driving systems.